%% file: main.tex
\documentclass[sigconf,screen]{acmart}
\usepackage{xspace}
\usepackage{longtable}
\usepackage{array}
\usepackage[table]{xcolor}

\usepackage{color}
\usepackage{nameref}
\usepackage{multirow}
\usepackage{tabularx}
\usepackage{censor}
\usepackage{nameref}
\usepackage{enumitem}

\AtBeginDocument{%
  }

\setcopyright{acmlicensed}
\copyrightyear{2018}
\acmYear{2018}
\acmDOI{XXXXXXX.XXXXXXX}
\acmBooktitle{Companion Proceedings of the 34th ACM Symposium on the Foundations of Software Engineering (FSE ’26), July 5 - 9, 2026 Montreal, Canada}
\acmISBN{978-1-4503-XXXX-X/2018/06}

\begin{document}

\title{Large Language Models in Game Development: Implications for Gameplay, Playability, and Player Experience}


\author{Keeryn Johnson}
\email{keeryn.johnson@ucalgary.ca}
\affiliation{%
  \institution{University of Calgary}
  \city{Calgary}
  \state{Alberta}
  \country{Canada}
}

\author{Muhammad Ahmed}
\email{muhammad.ahmed3@ucalgary.ca}
\affiliation{%
  \institution{University of Calgary}
  \city{Calgary}
  \state{Alberta}
  \country{Canada}
}

\author{Charlie Lang}
\email{charlie.lang@ucalgary.ca}
\affiliation{%
  \institution{University of Calgary}
  \city{Calgary}
  \state{Alberta}
  \country{Canada}
}

\author{Sahib Thethi}
\email{sahib.thethi@ucalgary.ca}
\affiliation{%
  \institution{University of Calgary}
  \city{Calgary}
  \state{Alberta}
  \country{Canada}
}

\author{Wilson Zheng}
\email{wilson.zheng@ucalgary.ca}
\affiliation{%
  \institution{University of Calgary}
  \city{Calgary}
  \state{Alberta}
  \country{Canada}
}

\author{Ronnie de Souza Santos}
\email{ronnie.desouzasantos@ucalgary.ca}
\affiliation{%
  \institution{University of Calgary}
  \city{Calgary}
  \state{Alberta}
  \country{Canada}
}

\begin{abstract}
\input{abstract}
\end{abstract}




\maketitle

\input{introduction}
\input{method}
\input{results}
\input{discussion}
\input{conclusion}

\nocite{*}

\bibliographystyle{ACM-Reference-Format}
\bibliography{bibliography}

\appendix

\end{document}

%% file: abstract.tex
This paper investigates how the integration of large language models influences gameplay, playability, and player experience in game development. We report a collaborative autoethnographic study of two game projects in which LLMs were embedded as architectural components. Reflective narratives and development artifacts were analyzed using gameplay, playability, and player experience as guiding constructs. The findings suggest that LLM integration increases variability and personalization while introducing challenges related to correctness, difficulty calibration, and structural coherence across these concepts. The study provides preliminary empirical insight into how generative AI integration reshapes established game constructs and introduces new architectural and quality considerations within game engineering practice.

%% file: introduction.tex
\section{Introduction}
\label{sec:introduction}

Over the years, game development has been investigated within the software engineering literature, including discussions of processes, challenges, requirements engineering, testing, and quality assurance in the context of video games \cite{kanode2009software, ampatzoglou2010software, santos2018computer}. The game software engineering research report sustained attention to requirements, usability, architecture, testing, and process management \cite{ampatzoglou2010software}. However, unlike general software systems that are typically evaluated in terms of functional correctness and usability, video games are described as interactive systems whose main aim is to provide fun and entertainment \cite{gonzalez2009usability}. In this context, general software characteristics alone are considered insufficient to achieve optimal player experience, indicating that game development must address additional concerns beyond traditional software projects \cite{gonzalez2009usability}. 

Within this context, gameplay, playability, and player experience emerge as key constructs \cite{fabricatore2007gameplay, gonzalez2009usability, sanchez2009playability, sanchez2012playability}. Gameplay has been defined as the set of activities that can be performed by the player and by other entities in response to player actions within the virtual world \cite{fabricatore2007gameplay}. Playability has been conceptualized as the instantiation of usability in games and as a construct that should be assessed throughout the development process to support player experience, with structured facets and measurable attributes \cite{fabricatore2007gameplay, gonzalez2009usability, sanchez2009playability, sanchez2012playability}. Player experience is related to engagement, enjoyment, and the interactive processes that occur during gameplay \cite{marshall2013games, mayra2011fundamental}. In this sense, gameplay concerns the activities available in the game, playability concerns the quality with which these activities can be performed, and player experience concerns the perceptions and responses that emerge during interaction \cite{fabricatore2007gameplay, gonzalez2009usability, sanchez2009playability, sanchez2012playability, marshall2013games, mayra2011fundamental}.

Recently, advances in artificial intelligence, particularly large language models, have introduced new forms of content generation, adaptive agents, and mixed initiative interaction into games \cite{sweetser2024large, yang2024gpt, pilaniwala2024future}. Large language models have been applied to procedural content generation, dialogue systems, gameplay-related design tasks, and mixed initiative development support \cite{sweetser2024large, yang2024gpt, pilaniwala2024future}. These studies report opportunities for increased variability and personalization, while also identifying challenges related to unpredictability, coherence, and evaluation \cite{sweetser2024large, yang2024gpt, pilaniwala2024future}. In this context, current research primarily documents applications and technical capabilities of generative models in game contexts \cite{sweetser2024large, yang2024gpt}. However, comparatively less attention has been given to exploring how the integration of large language models during development relates to established constructs such as gameplay, playability, and player experience.

Based on this problem, in this paper, we adopt an autoethnographic approach \cite{sharp2016role, zhang2019ethnographic, ellis2011autoethnography} to describe our experience developing two games that integrate large language models as part of their structure, with the goal of analyzing how the use of LLMs affected gameplay, playability, and player experience. Therefore, we are following this research question: \textbf{RQ.\textit{What is the impact of large language models on gameplay, playability, and player experience in the context of game development?}} By answering this question, we aim to provide preliminary empirical insights into how LLM integration influences gameplay structures, playability considerations, and player experience during development.

%% file: method.tex
\section{Method}
\label{sec:method}

This study adopts a collaborative autoethnographic approach to investigate how the integration of LLMs influenced gameplay, playability, and player experience in two game projects. Autoethnography is an empirical method that describes and systematically analyzes personal experience in order to understand broader practices \cite{ellis2011autoethnography, chang2016collaborative}. In contrast to traditional ethnography, where an external researcher observes a social setting, collaborative autoethnography positions members of the setting as co-authors who reflect on and interpret their own practice \cite{chang2016collaborative}. In this study, student developers document and analyze their lived experience integrating LLMs within a game software engineering context. Consistent with analytic expectations in autoethnographic research, narratives are not presented solely as personal accounts but are examined through a structured conceptual lens \cite{ellis2011autoethnography}. This work is situated within ethnographic traditions in software engineering that emphasize the analysis of socio-technical development practices and contextualized design processes \cite{zhang2019ethnographic, sharp2016role}. \\

\noindent\textbf{Context and Setting.}
The study was conducted in two educational settings: a third-year undergraduate course on software architecture and an undergraduate research project focused on software sustainability. Across these contexts, seven students developed two digital games over periods of three months and six months, respectively. Integration of LLMs was a mandatory component of both projects. Students were required to incorporate generative models into the game architecture as functional elements rather than auxiliary tools. Implemented uses included dialogue generation, narrative branching, non-player character behavior, and procedural content generation. Google Gemini and OpenAI models were used to support these functionalities. Development activities were embedded within standard software engineering workflows, including architectural design, implementation, iterative refinement, and testing. \\

\noindent\textbf{Participants and Authorship.}
Seven students participated in the projects, and five are co-authors of this paper. All teams contributed reflective narratives. Reflections were written individually and focused on development decisions, implementation challenges, and perceived effects of LLM integration on gameplay structures, playability characteristics, and player experience. Authorship reflects both direct involvement in development and intellectual contribution to the interpretation of experiences. The instructor acted as supervisor and co-author but did not include personal experience as primary data. \\

\begin{figure}[h]
\centering
\includegraphics[width=0.8\linewidth]{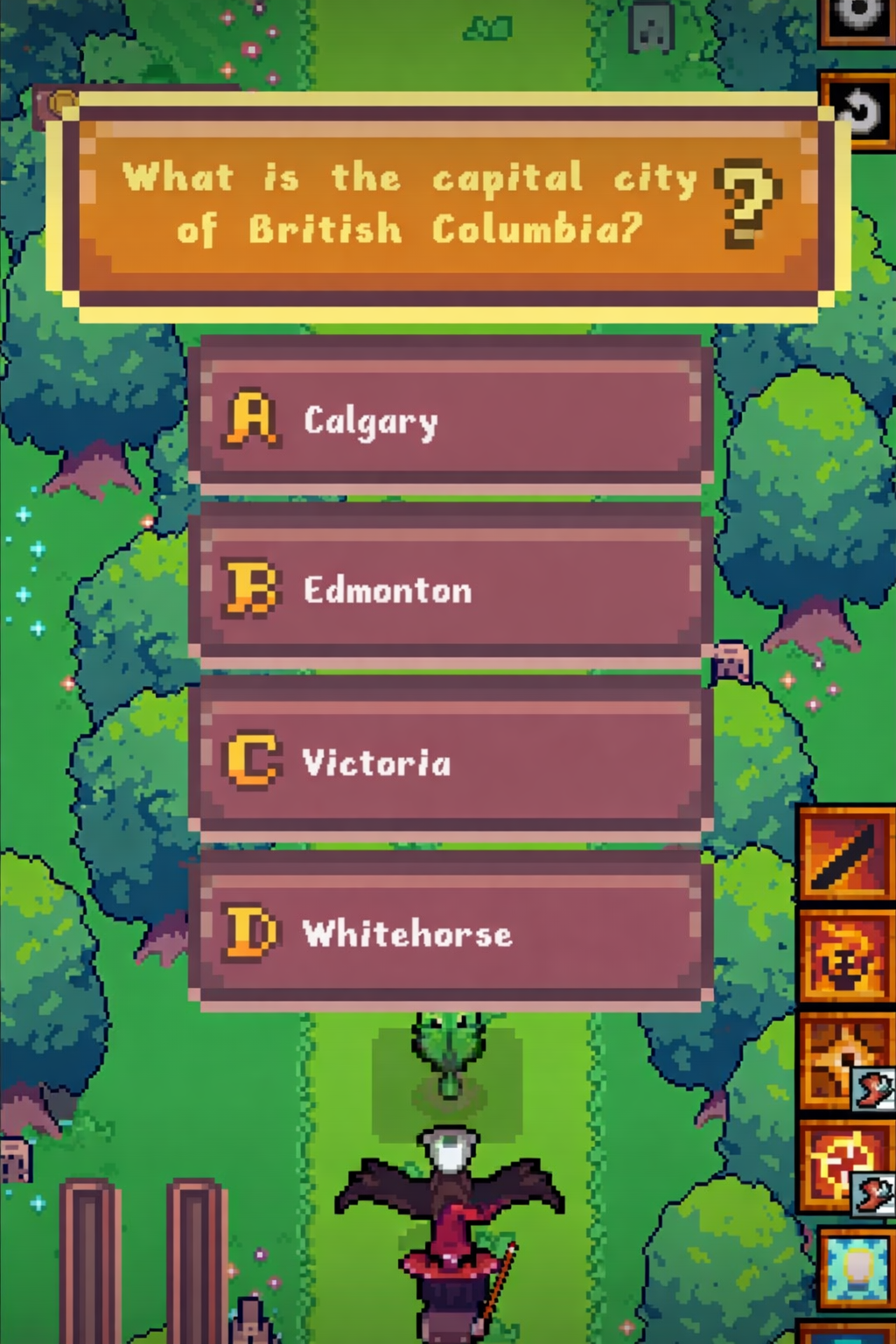}
\caption{WizdomRun game interface in use}
\label{fig:wizz}
\end{figure}

\begin{figure*}[h]
\centering
\includegraphics[width=0.9\linewidth]{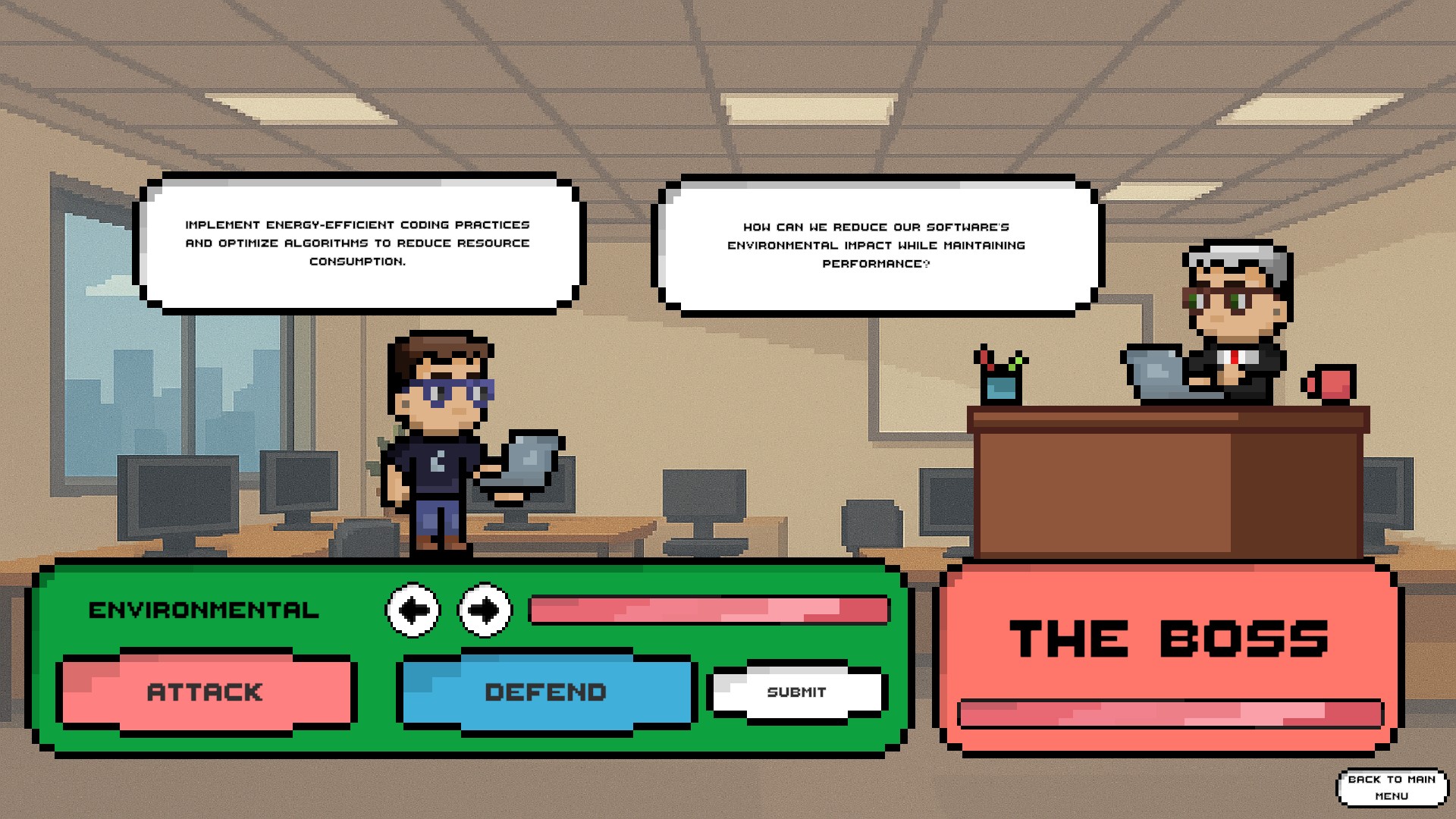}
\caption{Sena game interface in use}
\label{fig:sena}
\end{figure*}

\noindent\textbf{Game Artifacts.}
Two games, \textit{Wizdom Run} and \textit{Sena}, were developed as the empirical context of this study. In both cases, large language models were integrated as functional components that directly shaped gameplay structure and player interaction. \textit{Wizdom Run} (Figure \ref{fig:wizz}) is a study-oriented role-playing game implemented in Unity, structured around an auto-run campaign combined with combat mechanics. Players authenticate, create a character, and import personal study notes to initiate a campaign. The LLM processes imported documents using PDF extraction and generates multiple-choice questions across three difficulty levels. These questions are embedded into the core gameplay loop. During standard combat, correct answers replenish mana required for spell casting, while incorrect answers restrict available actions. Boss battles introduce a turn-based card system in which answering questions correctly grants bonuses or abilities. Questions are generated through OpenAI models, structured into JSON format, and stored in a PostgreSQL backend to support persistence of campaign state and player progress. In this game, the LLM functions primarily as a dynamic content generator that personalizes progression and links knowledge assessment directly to combat mechanics. \textit{Sena} (Figure \ref{fig:sena}) integrates LLM-driven interaction within a staged learning and gameplay structure focused on sustainability in software engineering. The system combines structured instructional content, conversational interaction, quiz-based reinforcement, and a scenario-driven game mode. The LLM supports dialogue-based clarification, scenario interpretation, and feedback generation. In the game mode, players evaluate sustainability-related decisions, select professional roles, and observe cumulative consequences represented through changes in the game state. Unlike \textit{Wizdom Run}, where the LLM primarily generates assessment questions tied to combat, \textit{Sena} uses the LLM to support reflective dialogue and consequence-oriented reasoning within a structured decision-making environment. Across both artifacts, LLMs were embedded as architectural components rather than auxiliary tools. Their integration influenced content variability, pacing, difficulty progression, and the coupling between player decisions and system responses, forming the basis for the experiential analyses in this study. \\

\noindent\textbf{Data Collection.}
The primary dataset consists of five reflective narratives, each approximately two pages in length. Reflections were written both during development and after project completion, allowing participants to document evolving perceptions of LLM integration. Students were guided by prompts asking them to reflect explicitly on gameplay, playability, and player experience. In addition to narratives, design documents and code repositories were consulted to contextualize implementation decisions and support triangulation of reported experiences. \\

\noindent\textbf{Analysis.}
Analysis combined thematic reading and content analysis \cite{drisko2016content}. First, narratives were read holistically to identify recurring themes related to the integration of LLMs. Second, segments of text were examined in relation to the constructs of gameplay, playability, and player experience, which functioned as sensitizing concepts guiding interpretation. Themes emerging across individual narratives were synthesized to characterize how LLM integration influenced structural aspects of gameplay, stability and coherence of playability, and perceived qualities of player experience. Experiences from both games were merged to identify common patterns rather than being treated as separate case comparisons. Analysis was conducted primarily by the instructor-researcher, with peer review and interpretive discussion involving co-authors to refine thematic interpretations. \\

\noindent\textbf{Reflexivity.}
Given the collaborative autoethnographic design \cite{chang2016collaborative, ellis2011autoethnography}, reflexivity was maintained throughout the study. Although the instructor served as course leader and supervisor, only student-authored narratives were treated as primary data. Triangulation across multiple individual accounts and consultation of development artifacts supported consistency of interpretation. Peer review among co-authors functioned as an additional mechanism to interrogate assumptions and refine themes. \\

\noindent\textbf{Threats to Validity.}
This study is context-bound and does not aim for statistical generalizability, consistent with collaborative autoethnographic traditions \cite{ellis2011autoethnography}. Findings are derived from a small number of student-authored reflections within two undergraduate development settings, which limits external validity. Insider positionality and reliance on self-reported narratives introduce potential interpretive bias and selective recall \cite{zhang2019ethnographic}. These threats were mitigated through triangulation across multiple individual reflections, consultation of design and code artifacts, and peer discussion among co-authors during analysis. Additionally, the study focuses on developer perspectives rather than independent player evaluation; therefore, conclusions regarding gameplay, playability, and player experience reflect development experience rather than external user testing. \\

\noindent\textbf{Ethical Considerations.}
All student participants contributed voluntarily and are co-authors of this manuscript. The reflections were written by the individuals represented in the study and are published with their explicit consent. In accordance with institutional guidelines, formal ethics approval was not required because the data consist of author-generated narratives. Participation was not linked to grading, and contributors retained the right to withdraw from publication.

%% file: results.tex
\section{Results}
\label{sec:results}

This section synthesizes participant reflections on how LLM integration influenced gameplay, playability, and player experience. \\

\noindent\textbf{Impact on Gameplay Structures.} LLM integration altered gameplay by coupling generated game elements directly to rule execution and progression systems. Rather than functioning as optional narrative additions, generated content and adaptive elements became operational components of advancement, resource management, and player actions. One participant emphasized that question generation in the game \textit{“directly drives skill point replenishment, boss encounters, and level progression, making it a core part of the gameplay”} (P1). In this configuration, generative outputs determined access to resources and continuation of core mechanics, positioning the LLM as an embedded rule-enforcing component within the game system. At the same time, LLMs introduced structured variability across repeated playthroughs by dynamically generating content tied to player-provided inputs. Differences in input materials altered the set of in-game elements, pacing, and exposure to challenge while the overarching mechanics remained stable. As one participant noted, \textit{“Each campaign offers a vastly different user experience, despite the underlying code remaining basic”} (P1). Another observed that \textit{“each playthrough will differ in its questions and answers”} (P2). However, this variability was bounded by both model configuration and system constraints. One developer reflected that when the same document was reused, \textit{“you generally get the same kinds of questions”} (P5), indicating that generative diversity was moderated by architectural decisions. In interactive contexts, player decisions were incorporated into subsequent system responses, such that \textit{“the LLM would be fed the decision that was made… allowing the system to generate a custom response”} (P4). In general, these accounts show that LLMs affected gameplay by layering probabilistic content generation onto stable mechanical frameworks, expanding possible game states without replacing deterministic rule structures. \\

\noindent\textbf{Impact on Playability.} While generative mechanisms expanded content flexibility, they simultaneously required architectural containment to preserve system coherence and functional reliability. Participants described the need to constrain outputs to predictable formats to ensure seamless integration with backend systems and gameplay logic. One participant explained that \textit{“ensuring that LLM responses were formatted exactly as expected… allowed for the backend design to remain cohesive and well-structured”} (P1). Another noted that explicitly defining the expected return format \textit{“allowed me to create objects in the C\# aspect of the game”} (P4), ensuring reliable parsing and system stability. These reflections indicate that maintaining playability depended on aligning generative outputs with structured schemas and validation processes. Challenge calibration also emerged as a tension. Although generated elements were intentionally categorized to support escalating difficulty (mentioned by P5), inconsistencies in how challenge levels were produced disrupted pacing. One participant observed that \textit{“some of the questions generated for medium or hard felt like the easy set”} (P3), challenging progression balance. Such mismatches required iterative adjustment to sustain coherent difficulty curves. The reliability of generated content further affected perceived gameplay fairness. Because player outcomes were directly tied to generated elements, incorrect outputs undermined legitimacy. One participant recalled that \textit{“a simple math question appeared… but none of the answer options were correct”} (P2). Another noted that incorrect responses \textit{“would tarnish the player's experience because they're not actually learning; they're just trying to figure out how the AI thinks”} (P5). Participants also reported patterned output structures, where the correct option \textit{“would show up in the same multiple-choice slot more than once”} (P3), potentially enabling unintended strategies. These accounts indicate that playability in LLM-integrated games is tightly coupled to output reliability, consistency, and controlled variability. \\

\noindent\textbf{Impact on Player Experience.} LLM integration enhanced perceived personalization by aligning generated content with player-provided inputs and contextual queries. Allowing users to introduce their own materials expanded the scope of in-game elements and individualized progression pathways. One participant remarked that \textit{“the range of topics that players can test themselves on is endless”} (P2). In conversational features, players could request clarification dynamically; as one developer noted, \textit{“the player could ask clarifying questions with additional context”} (P4). Such mechanisms were perceived as increasing engagement through contextual responsiveness and adaptability. However, experiential quality depended on balancing variability with predictability. One participant reflected that \textit{“having variable gameplay to a degree where there's still some expectancy offers more variety,”} while also emphasizing that \textit{“games need some level of predictability to remain fair and playable”} (P5). When contextual grounding was insufficient, the system could \textit{“simply answer like any publicly available LLM and didn’t determine if the question was on topic”} (P4), weakening coherence and trust. Incorrect outputs similarly disrupted immersion. These reflections suggest that player experience in generative game systems is shaped not only by personalization and dynamic content but also by perceived consistency.

%% file: discussion.tex
\section{Discussion}
\label{sec:discussion}

Existing studies on large language models in games primarily emphasize application domains such as procedural content generation, dialogue systems, and mixed initiative support, often framing variability and personalization as key benefits while noting challenges of unpredictability and evaluation \cite{sweetser2024large, yang2024gpt, pilaniwala2024future}. Our findings align with this body of work in that participants experienced increased variability and adaptive interaction as central effects of LLM integration. However, our study extends prior work by exploring how such integration reconfigures established game constructs at the architectural level. Rather than focusing solely on expressive potential, the findings show that coupling probabilistic generation to deterministic rule systems affects progression mechanics, difficulty calibration, and perceived gameplay fairness. In contrast to the literature that treats unpredictability primarily as a content-level concern \cite{sweetser2024large, yang2024gpt, pilaniwala2024future}, our findings indicate that variability operates as a structural property that must be mediated through prompt engineering, schema enforcement, and validation mechanisms.

When situated within software engineering research on game development, which emphasizes architecture, testing, requirements management, and quality assurance in game contexts \cite{kanode2009software, ampatzoglou2010software, santos2018computer}, our findings suggest that LLM integration introduces a distinct class of architectural and quality concerns. Traditional game software engineering highlights the importance of managing complexity and ensuring system quality through structured design and testing practices \cite{kanode2009software, ampatzoglou2010software}. In our study, variability becomes intertwined with rule consistency, progression control, and perceived gameplay fairness, shifting part of architectural control from fully pre-authored logic to generative subsystems. The need for prompt engineering, structured output schemas, and validation pipelines indicates that LLM components change coordination patterns within the architecture, requiring deliberate alignment between probabilistic generation and deterministic gameplay logic. The contribution of this study, therefore, lies in articulating how generative AI integration simultaneously expands gameplay variability and imposes new constraints on playability, quality assurance, and experiential trust within established software engineering practices for games.

Overall, our paper has implications for research and practice. From a research perspective, our findings suggest that future work should move beyond cataloguing generative applications and instead investigate how specific architectural integration strategies influence gameplay structure, playability stability, and player experience coherence. From a practice perspective, our findings indicate that effective LLM integration into games requires deliberate containment strategies embedded within standard software engineering workflows. Developers must treat correctness, formatting determinism, and difficulty alignment as gameplay-critical properties rather than peripheral implementation details. More broadly, our study suggests that integrating LLMs into games is not merely a matter of adding adaptive content but of redesigning the relationship between generative flexibility, structural control, and experiential reliability.

%% file: conclusion.tex
\section{Conclusions and Future Work}

This paper investigated how the integration of large language models influences gameplay, playability, and player experience within game development contexts through a collaborative autoethnographic study of two LLM-integrated games. Our findings indicate that LLM integration embeds generative mechanisms within core rule systems, increasing variability and personalization while introducing new constraints related to stability, correctness, and perceived gameplay. By coupling probabilistic generation with deterministic gameplay logic, LLM components require structured prompt design, validation, and calibration to preserve playability and coherence. Overall, these results provide preliminary empirical insight into how generative AI integration intersects with established game constructs and software engineering concerns. As future work, we will conduct user-centered empirical studies to extend this investigation beyond developer reflections and investigate how LLM-driven variability, correctness, and architectural coupling influence engagement and overall player experience during gameplay.

\section*{Data Availability}
Both games developed and analyzed in this study are open source. To preserve the integrity of the double blind review process, links to the corresponding public repositories will be provided in this section after the review of the manuscript. Sena is publicly available online: \url{https://www.senaplural.ca/}